\documentclass[manuscript]{aastex}

\def \ergsec{\hbox{erg$\,$s$^{-1}$}}

\def \phcmsec{\hbox{photons$\,$cm$^{-2}$s$^{-1}$}}

\def \gray {$\gamma$-ray }

\def \source {\hbox{0716+714~}}

\shorttitle{Extreme energetics by 0716+714.}

\shortauthors{Vittorini et al.2009}

\begin{document}

\title{ Powerful high energy emission of the remarkable BL Lac object S5~0716+714}
\vspace{2cm}
\author{ V. Vittorini\altaffilmark{1}, M. Tavani\altaffilmark{2},
A.~Paggi\altaffilmark{3}, A.~Cavaliere\altaffilmark{3},
A.~Bulgarelli\altaffilmark{5}, A.W.~Chen\altaffilmark{1},
F.~D'Ammando\altaffilmark{2,3}, I.~Donnarumma\altaffilmark{2},
A.~Giuliani\altaffilmark{6}, F.~Longo\altaffilmark{7},
L.~Pacciani\altaffilmark{2}, G.~Pucella\altaffilmark{2},
S.~Vercellone\altaffilmark{6}, A.~Ferrari\altaffilmark{1},
S.~Colafrancesco\altaffilmark{4}, P.~Giommi\altaffilmark{4}}

\altaffiltext{1}{CIFS--Torino, Viale Settimio Severo 3, I-10133,
Torino, Italy} \altaffiltext{2}{INAF/IASF--Roma, Via del Fosso del
Cavaliere 100, I-00133 Roma, Italy} \altaffiltext{3}{University of
Rome, ``Tor Vergata'', Via della Ricerca Scientifica 1, I-00133
Roma, Italy} \altaffiltext{4}{ASDC c/o ESRIN, Via G. Galilei snc,
I-00044 Frascati, Roma, Italy}
\altaffiltext{5}{INAF/IASF--Bologna, Via Gobetti 101, I-40129
Bologna, Italy} \altaffiltext{6}{INAF/IASF--Milano, Via E. Bassini
15, I-20133 Milano, Italy} \altaffiltext{7}{INFN--Trieste, Via
Valerio 2, I-34127 Trieste, Italy}
 \altaffiltext{*}{Email: \texttt{vittorini@roma2.infn.it}}

\begin{abstract}

BL Lac objects of the intermediate subclass (IBLs) are known to
emit a substantial fraction of their power in the energy range
0.1--10 GeV. Detecting \gray emission from such sources provides
therefore a direct probe of the emission mechanisms and of the
underlying powerhouse.

The AGILE \gray satellite detected the remarkable IBL
S5$\,$\source ($z \simeq 0.3$) during a high state in the period
from 2007 September - October, marked by two very intense flares
reaching peak fluxes of $200\times10^{-8}$ \phcmsec above 100 MeV,
with simultaneous optical and X-ray observations. We present here
a theoretical model for the two major flares and discuss the
overall energetics of the source.

We conclude that 0716+714 is among the brightest BL Lac's ever
detected at \(\gamma\)-ray energies. Because of its high power and
lack of signs for ongoing accretion or surrounding gas, the source
is an ideal candidate to test the maximal power extractable from a
rotating supermassive black hole via the \textit{pure}
Blandford-Znajek (BZ) mechanism. We find that during the 2007
\gray flares \source approached or just exceeded the upper limit
set by BZ for a black hole of mass $10^9\,$M$_{\odot}$.

\end{abstract}


\section {Introduction}

Blazars constitute a class of active galactic nuclei (AGN) that
often show very strong and rapid flux variability over the
electromagnetic spectrum. They are widely held to contain a black
hole (BH) with mass in the range $10^7 - 10^9\, M_{\odot}$, that
launches relativistic jets emitting highly non-thermal radiation.

The jet transports energy in electromagnetic form, and bulk plus
random kinetic energy of charged particles. The source radiation
may  also show a contribution by the accretion disk (including the
big blue bump, BBB), the broad line region (BLR), and a dusty
torus (see Urry \& Padovani 1995); lack or weakness of such
contributions mark out the class of BL Lac objects. The primary
energy of the jet may be supplied by power extracted from the
central rotating BH via interaction with its accretion disk
(Blandford \& Znajek 1977; Blandford \& Payne 1982; see also
discussions in Cavaliere \& D'Elia 2002 and McKinney 2005).

The blazar 0716+714 is a distant BL Lac at $z= 0.31\pm 0.08$
(Nilsson et al. 2008); its optical-UV continuum is so featureless
(Biermann et al. 1981; Stickel, Fried \& K$\ddot{u}$hr 1993) that
a redshift estimate has been possible only resolving and using the
host galaxy as a standard candle (Nilsson et al. 2008). This BL
Lac is of the IBL type, displaying a first broad peak in the
optical-UV bands and showing another broad peak near $1$ GeV; the
crossover between the two components falls in X-ray range (Massaro
et al. 2008; Ferrero et al. 2006). Recently, \source has been
detected by AGILE above 100 MeV several times during the period
September-October 2007 when the source was quite active in optical
band with variations on 1-day timescale (Villata et al. 2008). Two
bright \gray flares were detected. The first one reached a flux of
$\simeq(200\pm40)\times 10^{-8}$photons$\,$cm$^{-2}$s$^{-1}$ on
2007 September 11 with a photon index of $\simeq1.6$ and duration
$\le1\,$day. In 2007 October, following another prominent optical
activity detected by the WEBT consortium, AGILE and \emph{Swift}
satellites pointed again at the source. On 2007 October 23,
another 1-day bright flare was observed with a \gray flux
comparable to that detected in September (Chen et al. 2008).
Around this date a bright one-day flare and strong day-variability
were observed in the optical band (Villata et al. 2008); UV and
soft X-rays also showed strong and fast variability, whereas
modest or none variability appeared in the band 4-10 keV (Giommi
et al. 2008). Meanwhile, the radio flux showed a slow coherent
increase that remarkably begins around the day of the first \gray
flare and culminates around the date of the second \gray flare
(Villata et al. 2008). Thereafter \source was detected going back
to its ground state in all bands, with the \gray photon index that
softened toward $1.9$ (Chen et al. 2008) as previously observed by
EGRET (Lin et al. 1995).

In this paper, we present a physical model for the \source flaring
states, and study their extreme energetics with implications on
energy extraction from a rotating BH.

\section{Spectral modeling}

    \subsection{One Simple Model, Synchrotron Self-Compton }

The simplest homogeneous synchrotron self-Compton (SSC) model
assumes the blazar emissions to be produced in a ''blob'' of
radius $R$, containing relativistic electrons in a combination of
tangled and uniform magnetic field. The emitters move toward the
observer with bulk Lorentz factor $\Gamma$ (see, e.g., Tavecchio,
Maraschi \& Ghisellini 1998). We assume the emitters to emerge
from the injection/acceleration phase with a jet-frame
distribution of the random energies \(\gamma m c^2\) in the form
of a standard broken power-law
\begin{equation}
n_e(\gamma)=\frac{K\,\gamma_b^{-1}}{
(\gamma/\gamma_b)^{\zeta_{1}}+(\gamma/\gamma_b)^{\zeta_{2}}},
\end{equation}
where $\zeta_{1}$ and $\zeta_{2}$ are the spectral indices for
$\gamma<\gamma_b$ and $\gamma>\gamma_b$, respectively, $\gamma_b$
is the Lorentz factor at the break. These electrons emit a primary
synchrotron spectrum; a second contribution is then produced by
inverse Compton (IC) as the primary synchrotron photons scatter
off the same electron population. The spectral energy distribution
(SED) behaves as $\epsilon F(\epsilon)\propto \epsilon
^{1-\alpha}$, where $\epsilon$ is the energy of the received
photons, and $\alpha=(\zeta-1)/2$.

For electrons in a magnetic field $B$, the synchrotron SED peaks
around
\begin{equation}
\epsilon_s=h\,\frac{3.7\times10^6B\,\gamma_b^2\,\delta}{1+z}
\end{equation}
where $h$ is Planck's constant, $z$ is the redshift of the source,
and $\delta=[\Gamma(1-\beta\, cos\theta)]^{-1}$ is the bulk
Doppler factor due to the flow of emitters toward the observer at
an angle $\theta$ relative to the line of sight; the SED at the
synchrotron peak is
\begin{equation}
\epsilon_s\,F(\epsilon_s)\propto \delta^4R^3B^2K\,\gamma_b^2.
\end{equation}
As to the IC component, its SED contribution peaks at
\begin{equation}
\epsilon_c=\frac{4\gamma_b^2\epsilon_s}{3}
\end{equation}
with a peak value of
\begin{equation}
\epsilon_c\,F(\epsilon_c)\propto \delta^4R^4B^2K^2\gamma_b^4
\end{equation}
if the scattering takes place in the Thomson regime with the
density of target photons scaling as $n_{ph}\propto F_s R/c$. The
relativistic motion toward the observer amplifies the emitted
power by the factor $\delta^4$, and allows it to vary on a
timescale
\begin{equation}
t_{var}\gtrsim\frac{t_{cr}(1+z)}{\delta}
\end{equation}
close to or shorter than the crossing-time $t_{cr}=R/c$.

Due to the synchrotron and IC losses the electrons cool with
timescale
$$
\tau_{cool}(\gamma)=\frac{3mc}{4\beta^2\sigma_T\gamma(U_B+U_r)}
$$
where $\sigma_T$ is the Thomson cross section, $U_B=B^2/8\pi$ and
$U_r$ is the energy density of radiation before scattering. This
sets a typical cooling break at $\gamma_{cool}=3mc^2 / 4\sigma_T\,
R\,\beta^2(U_B+U_r)$ beyond which the electrons cool rapidly. In
the following, we take into account this constraint on the
particle distributions.

\textit{Simultaneous} multi-frequency observations  can provide
the five quantities $R,\,\,\delta ,\,\,B,\,\,K_e$ and $\gamma_b$
with the five Equations (2)-(6).

For BL Lacs with high frequency peaks (HBLs) requiring electrons
of higher energies ($\gamma_b>10^4$) the scattering  approaches
the Klein-Nishina (KN) regime with a blob-frame photon energy
$>m_ec^2/\gamma_b$. In the extreme KN regime the IC SED peaks at
$\epsilon_c\sim\gamma_b m_ec^2\delta/(1+z)$; the dependence on $B$
and $\gamma_b$ progressively weakens as the two latter parameters
grow.

\subsection{An Addition to the Model, External Seed Photons}
Additional target photon can be provided by a source external to
the jet (see Dermer et al. 2009). In this case, the high-energy
component of the spectra is due to the electrons that
Compton-scatter the external photons (EC); the SED now peaks at
energies
\begin{equation}
\epsilon_c=\frac{4\gamma_b^2\epsilon_{ext}'\delta}{3(1+z)}\, ,
\end{equation}
and the corresponding SED value is
\begin{equation}
\epsilon_c\,F(\epsilon_c)\propto
\delta^4R^3K\,\gamma_b^2N_{ext}'\epsilon_{ext}'\,.
\end{equation}

In this EC process two new ingredients enter: $\epsilon_{ext}'$
and $N_{ext}'$, respectively, the energy and the density at peak
of the external photons as seen by the moving blob. This has two
main consequences:\\
\textbf{i}) The model contains two further degrees of freedom and
the parameter evaluation may be degenerate;\\
\textbf{ii}) These news quantities are related to $N_{ext}$ and
$\epsilon_{ext}$ in the observer frame by means of the bulk
Lorentz factor $\Gamma$ in a manner that depends on the geometry
of the system (Dermer \& Schlickeiser 2002), causing an additional
dependence on $\Gamma$ in the EC spectra. Dermer \& Schlickeiser
(1993) discuss SED dependences on $\Gamma$ varying from
$\propto\Gamma^3$ to $\propto\Gamma^6$, for photons entering into
the blob from behind or head-on, respectively.

\subsection{Flux Variation Patterns}

Equations. (2) - (8) show that, in the synchrotron-IC framework,
variabilities of the first and second peaks are correlated,
possibly with a lag $t_{del}\sim t_{cr}(1+z)/\delta$.

When the energy fluxes  around these peaks $\phi\propto\epsilon
F(\epsilon)$ are  simultaneously monitored, we can compare the
corresponding light curves $\phi(t)$ in the respective band energy
$\epsilon_s$ and $\epsilon_c$, possibly with a time-lag $t_{del}$.
Then given two times $t_1$ and $t_2$, the ratio $r_{c}\equiv \phi
_{c} (t_2 + t_{del})/\phi _{c} (t_1 + t_{del})$ shows a dependence
on $r_{s} \equiv \phi _{s}(t_2)/\phi _{s}(t_1)$ that is related to
the emission mechanism. Here we consider some relevant cases.

If the variability is mainly due to electron
injection/acceleration, $K$ and/or $\gamma_b$ changes; then,
$r_c=r_s^2$ results if SSC dominates the IC emission. If instead
EC dominates then $r_c=r_s$ applies (compare Equation (3) with
Equations (5) and (8)). If variations are mainly due to changes in
$B$, then $r_c=r_s$ holds for SSC, and $r_c=1$ applies for EC. If
instead $\Gamma$ varies, we have $r_c=r_s$ in the SSC; in the EC
framework we have different behaviors depending on the geometric
of the jet relative to the external radiation: $r_c=r_s^x$ with
$x$ ranging from 3/4 for seed photons entering from behind, to 3/2
for photons entering head-on the flow.

\begin{table}
\small \noindent\caption{Variability Patterns.}
\begin{tabular}{|c|c|c|c|c|}
  \hline
  \bf{Model} & $\bf{\Gamma}$ changes & $\bf{B}$ changes & $\bf{K}$ changes &
  $\bf{\gamma_b}$ changes\\
  \hline
  \bf{SSC Th.}  &$r_c=r_s$ & $r_c=r_s$ & $r_c=r_s\,^2$ & $r_c=r_s\,^2$\\
  \bf{EC Th.} & $r_c=r_s\,^x$ & $r_c=1$ & $r_c=r_s$ & $r_c=r_s$\\
  \hline
  \bf{SSC KN}  &$r_c=r_s$ & $r_c\rightarrow r_s\,^{0.5}$ & $r_c=r_s\,^2$ & $r_c\rightarrow 1$\\
  \bf{EC KN} & $r_c=r_s\,^x$ & $r_c=1$ & $r_c=r_s$ & $r_c\rightarrow 1$\\
  \hline
\end{tabular}
\end{table}
\bigskip
\normalsize

In Table 1 we report the relations for the Thomson and the extreme
KN regimes (see also Paggi et al. 2009). The cooling of electrons
acts as a variation of $K$ and $\gamma_b$. Hence, in a flare due
to electron injection/acceleration, the trajectories $r_c$ versus
$r_s$ remain unchanged by pure radiative cooling.

    \subsection{Modeling the \gray flaring states}

A 1 day time lag between the emission in optical and
\(\gamma\)-ray bands during the two flares is apparent from
Figures 1 and 3 of Chen et al. (2008). This constrains the
emitting region radius and the Doppler factor to $R\le
5\times10^{16}(\delta/20)\mbox{ cm}$; the duration of both flares
is 1 day or less, which argues for cooling times
$\tau_{cool}(\gamma_b)\sim R/c$.

Moreover, optical and gamma light curves around the two flare
dates show evidence that $r_c=r_s^2$ applies (see Figure 2 and its
caption): this argues for a SSC process in the Thomson regime, and
concurs with the lack of sign of external gas to rule out EC
process (see also Table 1).

\begin{table*}
\begin{center}
\caption{Model Parameters for the 2007 \gray Flares of \source.}
 \small \noindent
\begin{tabular}{|c|c|c|c|c|c|c|c|c|c|c|}
  \hline
  \bf{Date} & \bf{Comp.} & $\bf{\Gamma}$ & \bf{B(Gauss)} & \bf{R(cm)} & $\bf{K(cm^{-3})}$ &
  $\bf{\gamma_b}$ & $\bf{\gamma_{min}}$ & $\bf{\zeta_1}$ & $\bf{\zeta_2}$ & $\bf{\tau_{cool}
  (\gamma_b)/t_{cr}}$ \\
  \hline
  Sept 11 & c I & 10 & 0.4   & $3\times10^{16}$ & 3.5 & 3800 & 100 & 2 &4.5 & 1.3 \\
          & c II & 15 & 0.3 & $3.5\times10^{16}$ & 1.4 & $7000$ & $3500$ &2&5 & 1.0 \\
          &single& 15 & 0.3 & $3.5\times10^{16}$  & 1.8 & 7000 & 60 & 1.8 &5&1.0\\

  \hline
  Oct 23 & c I & 10 & 0.4   & $3\times10^{16}$ & 2.5 & $4000$ & 40 &2&4.5 &1.2 \\
         & c II & 15 & 0.3 & $3.5\times10^{16}$ & 1.5 & $6500$ & 1300 &2&5 &1.1 \\
         &single& 15 & 0.3 & $3.5\times10^{16}$  & 1.8 & 6500 & 30 & 1.8 & 5&1.1 \\
  \hline
\end{tabular}
\end{center}
\end{table*}
\normalsize
\bigskip
Previous radio monitoring by the Very Long Baseline Array
 (VLBA) telescope of \source showed the presence of more
super-luminal components relative to the active state of 2003-2004
(Rastorgueva et al. 2009). Moreover, the absence of the signatures
of IC catastrophe provided a lower limit $\delta\ge 14$ for the
Doppler factor (Ostorero et al. 2006; Fuhrmann et al. 2008; see
also Wagner et al. 1996); Bach et al. (2005) also argue for high
Doppler factors in these fast variable components and for a
viewing angle $\theta\le5^{o}$.

We note that modeling the SED of \source with a standard one-zone
SSC model would fail to reproduce the simultaneous radio, optical,
X-ray and \gray data for the October flare. In particular, the
crossover in the  X-ray band would not be well reproduced, and the
hard X-ray flux would be overestimated by a factor $\sim$3;
furthermore it would be difficult to model the hardness of the
\gray spectrum during the September flare (Figure 1 red dashed
lines). A one-component model also hardly explain together the
slow trends of the radio, optical and hard X-ray bands and the
faster variability observed in the optical, soft X-ray and \gray
bands (see Villata et al. 2008, Giommi et al. 2008, Chen et al.
2008). Hence, we adopt a two-component model: the first produces
the slowly variable radio and hard X-ray bands, whereas the second
is responsible for the faster variability in optical-UV, soft X-,
and \gray bands. Despite these problems, the possibility to fully
constrain by observations the parameters stimulate us to also show
for comparison a one-component model.

The SED of all these models are shown in Figure 1 and the
parameters are listed in Table 2; a viewing angle
\textbf{$\theta\approx 2^{o}$} is adopted according to Bach et al.
(2005).

\section{The extreme energetics of 0716+714}

Under the assumption of isotropic emission, the observed power
radiated from a source with luminosity distance  $D_L(z)$ is
$$L_{obs}=4\pi D_L(z)^2\int d\epsilon F(\epsilon)$$.

The jet transports a total power
$P_{tot,flare}=L_r\,+\,L_{kin}\,+\,L_B$ contributed by intrinsic
radiated power, kinetic energy flow of the electrons and of the
cold protons (with one proton per emitting electron), and Poynting
flux, respectively:
\begin{eqnarray}
\noindent
&L_{r}&=L_{obs}\Gamma^2/\delta^4=2\times10^{-3}(\delta/15)^{-4}{\Gamma_1}^2\,L_{obs},\\
&L_e&=0.8\times10^{43}R_{16}^2\,{\Gamma_1}^2\,{\langle{n_e\gamma}\rangle}_{4}\,\rm erg\,s^{-1},\\
&L_p&=1.4\times10^{43}R_{16}^2\,{\Gamma_1}^2\,{\langle{n_p}\rangle}_{1}\,\rm erg\,s^{-1},\\
&L_B&=3.8\times10^{43}R_{16}^2\,{\Gamma_1}^2B^2\,\rm erg\,s^{-1},
\end{eqnarray}
see also Celotti \& Ghisellini (2008). The latter authors show
that in BL Lacs \(L_r\) tends to match the sum of the other
contributions. In fact, for the two flares of \source at redshift
$z=0.31$ we find \(L_r \simeq 2\times{10}^{45}\mbox{ erg}
 \mbox{ s$^{-1}$}\), and from our two-component SSC model (with the
parameters listed in Table 2) we obtain a total jet power
$$
P_{tot, flare}= (3.5\pm 1)\times10^{45}\,\rm erg\,s^{-1}
$$
with \(L_r\gtrsim (L_e+L_p+L_B)\). Under this condition the total
jet power is minimized and the details of cooling does not affect
materially the global energetics, being the radiated luminosity
\(L_r\) mainly contributed by peaks emission. Moreover, the
uncertainty in \(P_{tot, flare}\) is mainly due to the observed
\gray flux error.

For the one-component model, we obtain $P_{tot, flare}= (1\pm
0.5)\times 10^{46}\,$\ergsec and \(L_r\lesssim (L_e+L_p+L_B)\)
holds. In this case, the total jet power is not dominated by the
radiated one, but the parameters now are well constrained by the
observation and the uncertainty is still due to the flux error.

\subsection{Testing the Blandford-Znajek Mechanism}

Here, we compare the jet powers provided by our models with the BZ
mechanism; this set a limit on the power extractable from a
rotating BH
\begin{equation}
P_{BZ}\simeq 2\times10^{45}M_9\,\rm erg\,s^{-1}
\end{equation}
under conservative values of \(B\), as discussed in Cavaliere \&
D'Elia (2002, and references therein; see also our discussion).

Estimating the redshift and the BH mass of \source is not trivial
because of the lack of emission lines. Recently, however, Nilsson
et al. (2008) pinpointed the host galaxy of this source and
reported a determination of its redshift at $z=0.31\pm 0.08$.
Then, for this host, a BH of mass $M\sim5\times10^8M_{\odot}$
should accord with the fundamental plane of BL Lacs (Falomo et al.
2003).

On the other hand, some authors using  micro-variability of the
optical flux estimate the mass of the central BH for \source by
\begin{equation}
M<\frac{c^3\tau\,\delta}{6G(1+z)}\approx
3\times10^8M_{\odot}\big{[}\frac{\delta}{20}\,\frac{\tau}{600}\big{]}.
\end{equation}
Considering $\tau\simeq 450\,$s as the shortest variability time
found by Sasada et al. (2008) one obtains
$M<2\times10^8(\delta/20)M_{\odot}$. Gupta, Srivastava \& Wiita
(2009) obtain the more stringent constraint
$M<5\times10^7$M$_{\odot}$ for the case of a Kerr BH. However,
these authors discuss that the micro-variability in \source may be
due to a small region in the jet or due to internal disk modes, so
that the BH mass would be higher.

We note that a BH of mass $M<10^8$M$_{\odot}$ in \source would
imply in Equation (13) a power limit $P_{BZ}<<L_r$ inconsistent
with the power $L_r$ emitted during the flares.

   \section{Discussion and Conclusions}

We modeled the \source flares of 2007 September 11 and October 23
with a two-component SSC model, similarly to Tavecchio $\&$
Ghisellini (2009). Despite the larger number of model parameters,
we believe that our two-component modeling reproduces the complex
variability and the hard \gray spectra of \source better than a
one-component model.
\begin{figure}
\vspace{7cm} \includegraphics{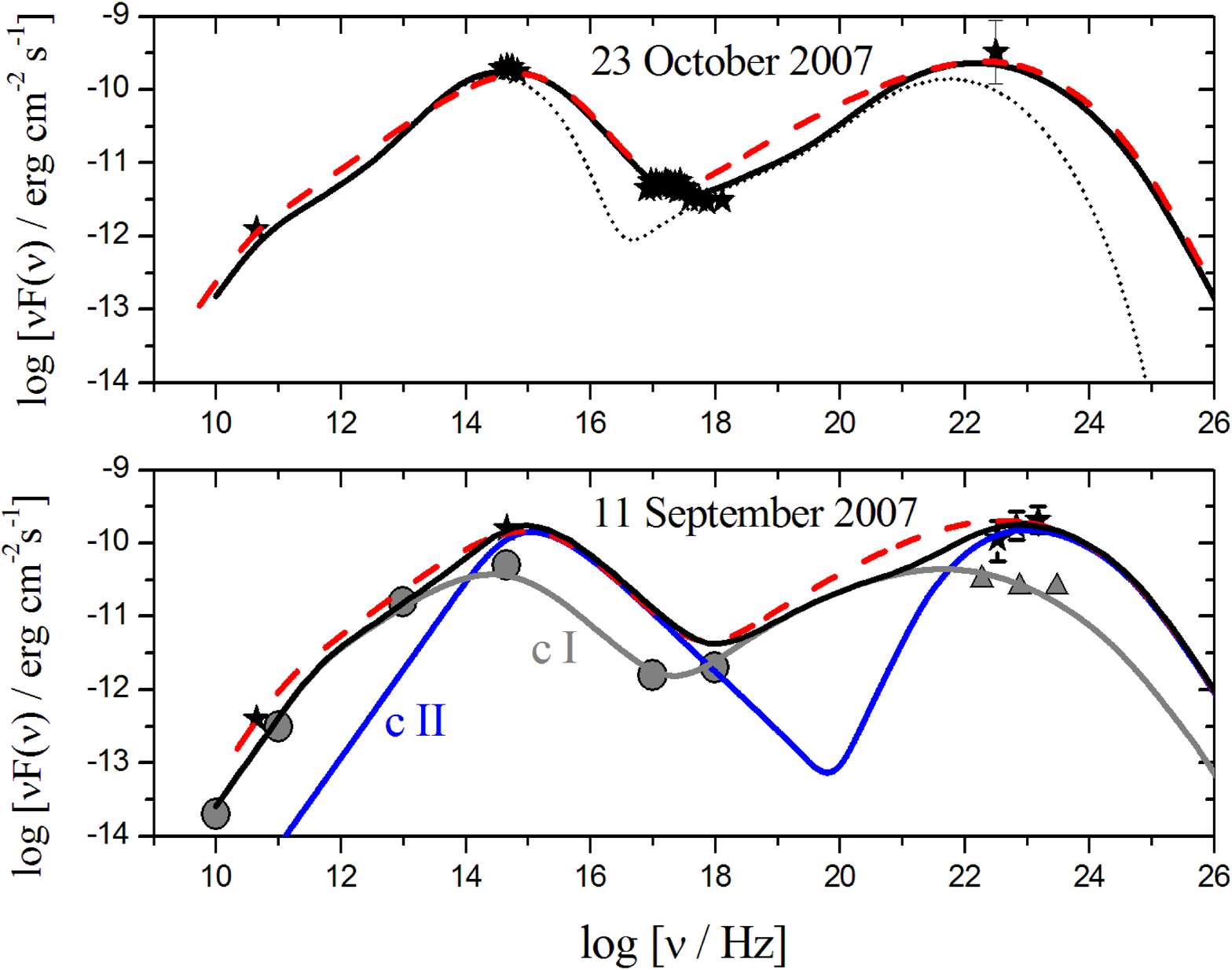} \vspace{2cm} \caption{ Bottom panel: SED
relative to the 2007 September 11 data. Curves labeled "c I" and
"c II" (gray and blue solid lines) show the two separate
components. Data in light gray represent a previous low state plus
the EGRET observation in \gray (see Lin et al. 1995). Top panel:
two-components SED relative to the 2007 October flare (solid
line), and after 1 day by radiative cooling (dotted line). In both
panels black stars are simultaneous data (see Villata et al. 2008;
Giommi et al. 2008; Chen et al. 2008), the black lines are the
two-components models and the dashed red lines represent the best
one-component models.}\vspace{0.5cm}
\end{figure}

\begin{figure}
\includegraphics{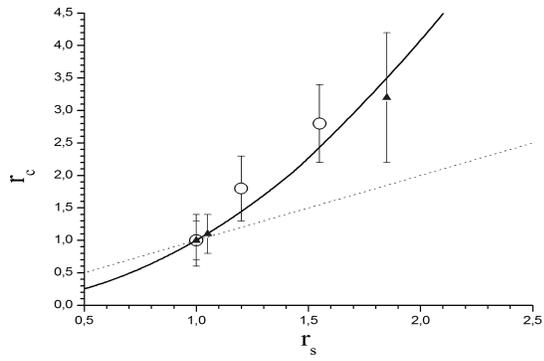} \vspace{10cm} \caption{Variability plane with the
trajectories of 0716+714 based on simultaneous optical and \gray
observations. Open circles are for 2007 September, and filled
triangles for 2007 October data. The thick solid line represents
the quadratic trend, expected from electrons injected or
accelerating in the SSC framework; for comparison, the linear
behavior expected from EC is represented by the thin dotted line.}
\end{figure}

Our Figure 2 indicates a quadratic dependence between the
synchrotron and IC fluxes. This concurs with the lack of emission
lines and BBB to rule out models involving external sources of
seed photons that produce a linear dependence, and to strongly
support electrons radiating in the Thomson regime within the SSC
framework as the most likely radiation process. We refer to our
Figures 2 and 1 in Chen et al. (2008) to show that the 2007
September data lie around the maximum; in time, the trajectory
starts from the lower $r_c$ value, attains a maximum and then
falls down. In 2007 October, the trajectory describes the decline
of a flare starting from the higher $r_c$ value.  Rises and falls
both occur on 1-day timescale, and the trajectories are quadratic.
This suggests the electron acceleration and cooling to occur on
similar timescales.

\begin{figure}
\vspace{7cm} \includegraphics{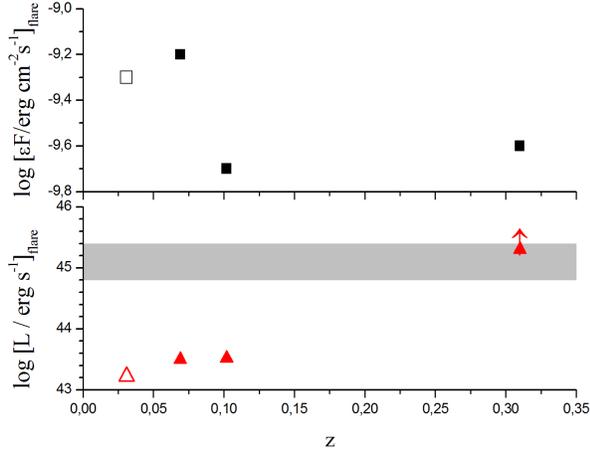} \vspace{2cm} \caption{Observed high-energy
peak fluxes (top panel) and the corresponding intrinsic peak
luminosities (bottom panel) for a number of BL Lac objects during
their flaring states, ordered with their $z$: Mrk 421 (open
symbols, here shown for reference), BL Lac, W Comae, and \source.
For the first three objects, data are respectively from Donnarumma
et al. (2009), Ravasio et al. (2002), and B\"ottcher et al.
(2002). The shaded area represents the BZ limiting luminosity
range for a BH mass in the range $(3\times 10^8\, -\, 10^9) \,
M_{\odot}$. The arrow points at the total jet power by also adding
the  components of  Equations (10) - (12).}
\end{figure}

The crossover between the synchrotron and IC branches of the SED
is located in X-rays (see also Foschini et al. 2006; Ferrero et
al. 2006), and is contributed by both the slowly variable
component (I) and the faster component (II).

As to the faster components (II) adopted for the two flares, they
are marked by high electron energies $\gamma_b\sim 7\times10^3$
with a sharp low energy cutoff $\gamma_{min}\sim2\,10^3$ (see also
Tsang \& Kirk 2007). Moreover, high bulk Lorentz factor
$\Gamma=15$ (that is $\delta\simeq23$) is used in accord with
Wagner et al. (1996). In Table 2, it is shown that our choice of
parameters implies $\tau_{cool}(\gamma_b)\simeq t_{cr}$ for the
\textit{fast} components II: those quench very rapidly causing
strong variability in the optical-UV, soft X- and \gray bands.
Little or no variability results in radio and hard-X ray bands
produced by the rising part of the \textit{slow} components I.
This behavior is in agreement with the complex multi-band
variability reported by Chen et al. (2008), Giommi et al. (2008),
and Villata et al. (2008).

Considering the redshift \(z= 0.31\) of \source, we find that its
intrinsic radiative luminosity is of order \(L_r\approx
2\times{10}^{45}\mbox{ erg} \mbox{ s$^{-1}$}\). On adding the
other jet components in Equations (10) - (12), the total power
becomes \(P_{tot, flare}\approx 4\times{10}^{45}\mbox{ erg} \mbox{
s$^{-1}$}\) for the two-component model,  and \(P_{tot,
flare}\approx 2\times{10}^{46}\mbox{ erg} \mbox{ s$^{-1}$}\) for
the one-component model: the source exceeds the BZ limit
\(P_{BZ}\approx 2\times{10}^{45} M_{BH}/M_9\mbox{ erg} \mbox{
s$^{-1}$}\) (see Figure 3). This obtains for a maximally rotating
BH of $10^9\rm M_{\odot}$(spun by past accretion) via interaction
with a disk magnetic field \(B\sim {10}^4\mbox{ G}\) sustained by
gas or radiation pressure, in the absence of ongoing accretion
(see discussion by Cavaliere \& D'Elia 2002). The
\textit{simultaneous} nature of our multifrequency data during the
2007 \gray flares of \source is a crucial ingredient of our
result. Other BL Lacs with weak or negligible accretion disk or
BLR contributions have been reported to attain (model dependent)
high total luminosities, e.g., 2032+107 with an inferred
$P\sim10^{46}\rm erg\,s^{-1}$ (see Celotti \& Ghisellini 2008).
Corbel \& Reyes (2008, ATel 1744) report the high \gray peak flux
recently attained by 0235+164 at $z\simeq0.94$  for which,
however, EC contributions cannot be ruled out. However, to our
knowledge, only the simultaneous data obtained by our group for
\source provide a model independent evaluation of the total jet
power from a BL Lac being close or just above the BZ limit.

We also show in Figure 3 the intrinsic radiated power for other BL
Lacs with intense \gray and TeV emissions as for the 1997 flare of
BL Lacertae with peak flux around $500\times10^{-8}\,\phcmsec$ but
with lower redshift $z=0.069$ (Bloom et al. 1997; see also Ravasio
et al. 2002) and intermittent evidence of lines and thermal
emissions. We also report W Comae with redshift $z=0.102$ but
\gray flux at levels of $50\times10^{-8}\,\phcmsec$ (B\"ottcher,
Mukherjee \& Reimer 2002).

The increasing trend of $L_{r}$ with \(z\) is likely to arise from
the Malmquist bias, while the decrease at low \(z\) may result
from sampling a limited cosmological volume. Nevertheless, we
stress that up to now no BL Lac source has sharply exceeded the BZ
limiting power. It will be worthwhile to keep under close watch
the most distant BL Lacs (despite the obvious monitoring
difficulties) to catch powerful emissions. If violations will be
found, these may be possibly discussed in terms of the
Blandford-Payne mechanism (Blandford \& Payne 1982) that, however,
requires ongoing accretion not supported in the case of \source.
Alternatively, higher powers may be attained with higher magnetic
fields up to \(B^2/4\pi\lesssim\rho c^2\) related to the plunging
orbits (see Meier 1999), which however imply short source
lifetimes in the absence of accretion.

In conclusion we find the total power transported in the jet of
0716+714 to be $P_{tot, flare}\sim 4\times10^{45}\rm
$erg$\,$s$^{-1}$, and so to approach the limit of the BZ mechanism
for a BH up to $10^9$M$_{\odot}$ with \textit{conservative} values
of \(B\). Such high powers in \source constitute an unescapable
consequence of two observed facts:
\begin{itemize}
  \item The \gray flux attains high levels in the SED during the two flares, with
an emitted power comparable with the optical one as shown by
simultaneous observations (see Figure 1);

  \item The lack of external sources of seed photons related to
emission lines or BBB concurs with the observed quadratic
dependence $r_c=r_s^2$ to rule out EC contributions to the high
energy hump (see Figure 2) and considerable ongoing accretion.
\end{itemize}
\noindent {\bf Acknowledgments}: This investigation was carried
out with partial support under ASI contract no. I/089/06/2.

{}

\bibliographystyle{apj}

\end{document}